\documentclass[prl,twocolumn,showpacs,amsmath]{revtex4}
\usepackage{graphicx}
\newcommand\pictc[5]{\begin{figure}[t]
            \centerline{\vspace{0mm}
            \includegraphics[width=#1\columnwidth,height=0.7\textheight,keepaspectratio]{#3}}
            \protect\caption{\protect\label{fig:#4} #5}%\vspace{-2mm}
                    \end{figure}            }

\newcommand\pict[4][1.0]{\pictc{#1}{!tb}{#2}{#3}{#4}}
\newcommand\rpict[1]{\ref{fig:#1}}

\begin{document}

\title{Observation of attraction between dark solitons}

\author{Alexander Dreischuh$^{1,2,3}$, Dragomir N. Neshev$^1$, Dan~E. Petersen$^2$, Ole Bang$^4$, and Wieslaw Krolikowski$^2$}
\affiliation{$^1$Nonlinear Physics Centre, $^2$Laser Physics Centre, Research School of Physical Sciences and Engineering, The Australian National University, Canberra ACT 0200, Australia\\
$^3$Department of Quantum Electronics, Faculty of Physics, Sofia University, Sofia, Bulgaria\\
$^4$COM$\bullet$DTU Department of Communications, Optics \& Materials, Technical University of Denmark, DK-2800 Kongens Lyngby, Denmark}

%\author{Wieslaw Krolikowski}
%\affiliation{Laser Physics Center, Research School of Physical Sciences and Engineering, The Australian National University, Canberra ACT 0200, Australia}

\begin{abstract}
We demonstrate a dramatic change in the interaction forces between dark solitons in nonlocal nonlinear media. We present, what we believe is the first experimental evidence of attraction of dark solitons. Our results indicate that attraction should be observable in other nonlocal systems, such as Bose-Einstein condensates with repulsive long-range interparticle interaction.
\end{abstract}


\maketitle

It is commonly accepted that solitons, i.e. localized waves propagating without changing their shape, are ubiquitous in nature~\cite{Remoissenet} and are native to many diverse systems like plasmas, molecular chains, spin waves, atmospheric physics, superfluidity, nonlinear optics, and Bose-Einstein condensates (BECs). A soliton can form when the dispersive or/and diffraction processes associated with the finite size of the wave are counterbalanced by the wave self-induced change of the properties of the medium. In the context of BECs, for example, the soliton represents a coherent excitation of a matter wave~\cite{BEC_soliton}, whereas in optics it is a localized light beam or pulse~\cite{KivsharAgrawal}. The continuous interest in solitons is stimulated by their unique collisional properties, i.e. they behave like particles displaying forces in their mutual interaction. Furthermore, the fundamental features of their interaction are of rather universal character. Thus, for example, matter wave solitons interact basically in the same way as optical or plasma solitons.

There are two fundamental types of solitons: bright, in the form of a localized structure~\cite{Nail_book} and dark, in the form of a localized dip on a plane-wave background~\cite{Swartzlander91,spatial,temporal,Kivshar98,dark_BEC,Proukakis04}. Already early studies pointed out a distinctive difference between the interaction of bright and dark solitons. Bright solitons can either attract or repel depending on their relative phase.  The phase between dark solitons is  fixed as they are formed on a single background wave and {\em they are believed to always repel}~\cite{BlowDorran85,Zhao89,Swartzlander91,YKWK95}. This fact imposes a fundamental limit to the applicability of dark solitons and all existing soliton applications are currently based only on bright solitons.

Recently, we predicted theoretically~\cite{Nikola04} that the nature of dark soliton interaction may change drastically in the presence of a nonlocal response of the material. In this letter we report the first experimental observation of {\em attraction between spatial dark solitons} in any physical system. Our findings open new possibilities for control of the interaction between dark solitons, which we believe will revive the interest towards them and will allow for their broader applicability.

In nonlocal nonlinear media the nonlinear response induced at a certain point is carried away to the surrounding regions. In this way a narrow localized wave can induce a spatially broad response of the medium \cite{Snyder97}. Spatial nonlocality is an inherent property of many physical systems. It often results from transport processes, such as atomic diffusion~\cite{Suter} or heat conduction~\cite{thermal_nlc}. Spatial nonlocality is also natural for media with a long-range inter-particle interaction including, for instance, dipolar BECs~\cite{ParSalRea98}, or nematic liquid crystals with long-range molecular reorientational interactions~\cite{lc1}. It appears that nonlocality leads to novel phenomena of generic nature. For instance, it may promote modulational instability in self-defocusing media~\cite{Wyller,WK04}, as well as suppress wave collapse and stabilize multidimensional solitons in self-focusing media~\cite{Tur85,Bang02,df,dahli05}. Nonlocal nonlinearity may even describe parametric wave mixing~\cite{Nikola03}. Furthermore, nonlocality significantly changes bright soliton interaction~\cite{Peccianti}.
%Interactions between dark solitons are also expected to be affected by the nonlocality in the system.

%--------------------------------------------------------------------------
\pict{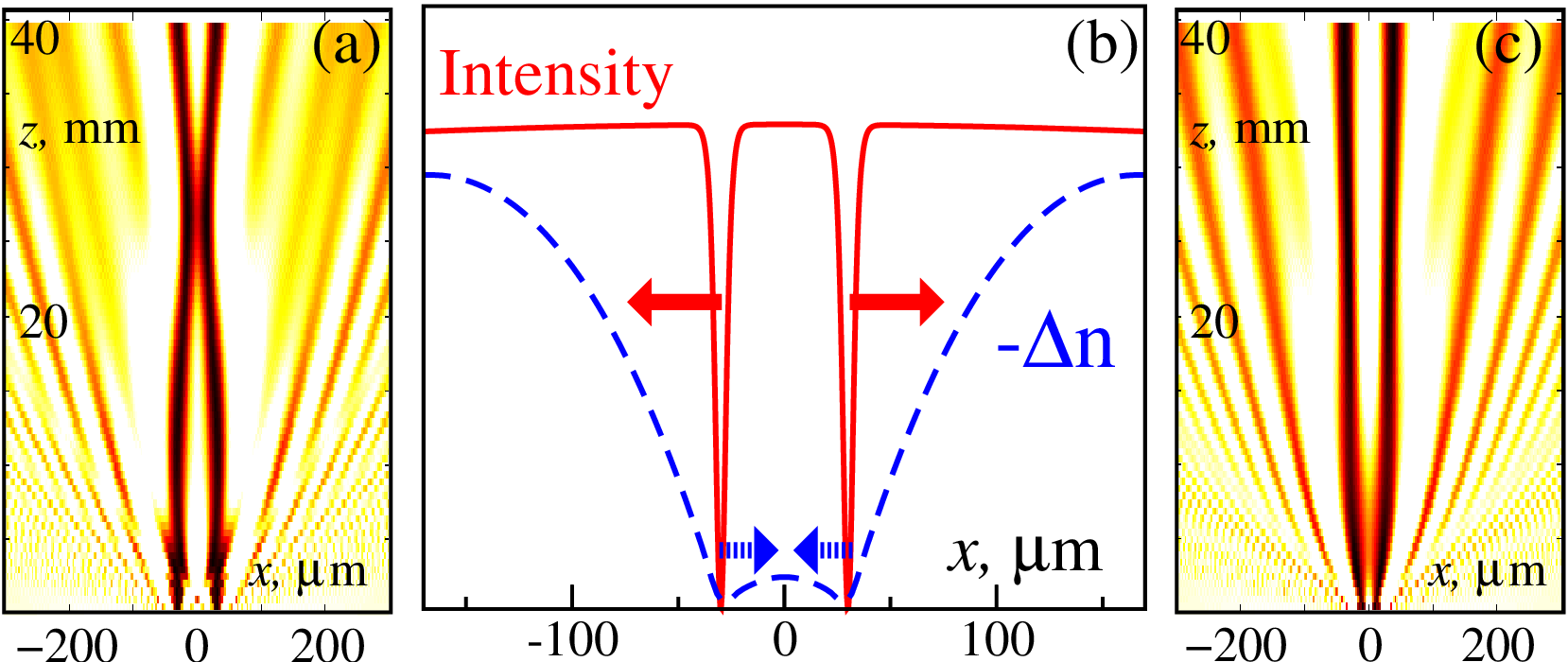}{fig1}{(color online) (a) Trajectories of two attracting dark solitons separated at 60$\mu$m. (b)~Initial intensity (solid line) and corresponding index structure (dashed line) generated by the nonlocal nonlinear response, giving rise to attractive (dashed arrows) and repulsive forces (solid arrows) on the solitons. (c)~Trajectories of two closely spaced dark solitons (20$\mu$m) with dominating repulsive interactions.}
%--------------------------------------------------------------------------

To test experimentally how the nonlocality affects the forces acting between dark solitons, we considered the propagation of an optical beam in a weakly absorbing liquid. Light absorption increases the temperature of the liquid and subsequently decreases its density and refractive index, resulting in a defocusing nonlinearity. In addition, heat conduction leads to a temperature and consequently refractive index profile much wider than the light beam itself, indicating the {\em inherently nonlocal} character of the thermal nonlinearity. The choice of the liquid medium provides additionally the advantage to monitor the beam profile along the entire propagation without destructive intervention to the material.
%a property not available in other physical systems.

The propagation of an optical beam along the $z$-axis of a weakly absorbing liquid
is described by the system of normalized equations~\cite{thermal_nlc} for the slowly varying field amplitude $\psi(x,z,t)$
\begin{eqnarray}
& & \frac{\partial \Delta n}{\partial t} - \frac{\partial^2 \Delta n}{\partial x^2}=-\alpha{|\psi|^2},
\label{heat} \\
& & i\frac{\partial\psi}{\partial z} + \frac{ \partial^2 \psi}{\partial x^2}
   + \Delta n(I)\psi +i{\alpha}\psi= 0.
\label{NLS}
\end{eqnarray}
Here the longitudinal $z$ and transverse $x$ coordinates are scaled by the wave vector $k=2\pi n_0/\lambda$ ($z\rightarrow kz/2, x\rightarrow kx$). The parameter $\alpha$ (scaled as $\alpha \rightarrow \alpha/k$) represents the linear absorption. Time in Eq.~(\ref{heat}) is normalized to $k^2D$, where $D$ is the thermal diffusivity of the medium, and $\lambda$ is the laser wavelength.
%and   $k=2\pi/\lambda n_0 $ is the wavenumber in the material. $\lambda$ is a wavelength while $n_0$ is the linear refractive index.
%, $\omega$ is the optical frequency,
Eq.~(\ref{heat}) is a heat equation describing the temporal and spatial dynamics of the refractive index change of the medium $\Delta n(I)$ induced by a heat source in the form of a beam with intensity $I(x,z,t)=|\psi(x,z,t)|^2$. Eq.~(\ref{NLS}) is a nonlinear Schr\"{o}dinger equation governing the evolution of the beam amplitude in the presence of nonlinear refractive index change. We solved these equations numerically using a split-step Fourier method for the propagation equation and a finite difference method for the heat equation. For simplicity we assumed unrestricted heat flow in the transverse direction. Using typical thermal parameters of mineral oils yields the diffusion coefficient $D=10^{-7} m^2/s$ and linear index $n_0=1.5$ for $\lambda=0.532\mu$m. The  absorption coefficient is  set to $\alpha=0.01cm^{-1}$ which results  in 10\% power loss over 50mm of propagation. As an initial condition we used a broad Gaussian beam of peak intensity $I_0$ and full width at half maximum (FWHM) 2.8~mm, with two closely placed $\pi$ phase jumps. Such initial conditions result in the formation of two ``black'' solitons with a zero individual transverse velocities~\cite{Zakharov}.

An example of steady-state trajectories of two adjacent solitons separated by a distance of 60$\mu$m is shown in Fig.~\rpict{fig1}(a). The inward bending of the trajectories clearly indicates the presence of attractive forces. For longer propagation the separation between the two solitons oscillates, leading to formation of an oscillatory-type bound state. This behaviour strongly differs from the interaction of dark solitons in local nonlinear media, where the soliton trajectories diverge due to repulsion.

The physics of this interaction can be intuitively explained as follows. The two close dark solitons with an intensity profile depicted in Fig.~\rpict{fig1}(b) always try to repel because of local refractive index drop in the overlapping region (repulsive forces are indicated as solid arrows). However, in a nonlocal medium these solitons induce also a large scale change in the refractive index in the form of a broad trapping potential [Fig.~\rpict{fig1}(b, dashed curve)]. This potential provides an attractive force (indicated by dashed arrows) which counteract the natural repulsion of the solitons. Ultimately, the interplay between these two forces will determine the outcome of the interaction. While repulsion prevails for close initial separations and the solitons diverge as depicted in Fig.~\rpict{fig1}(c), its strength decreases with increase of the separation, allowing the nonlocality mediated attraction to become dominant at larger separations.
%Hence, for a certain separation distance repulsive and attractive forces may cancel out providing a necessary condition for the formation of a bound state of solitons. Whether such a bound state will actually form depends also on soliton velocities. For initially close solitons their velocities acquired via repulsion may be high enough to overcome attraction so no trapping occurs and the solitons eventually escape as depicted in Fig.~\rpict{fig1}(c).
%This interaction scenario is completely analogous to the dynamics of a particle moving in an asymmetric potential field, and it lends further support to the concept of the particle-like nature of solitons.

%--------------------------------------------------------------------------
\pict{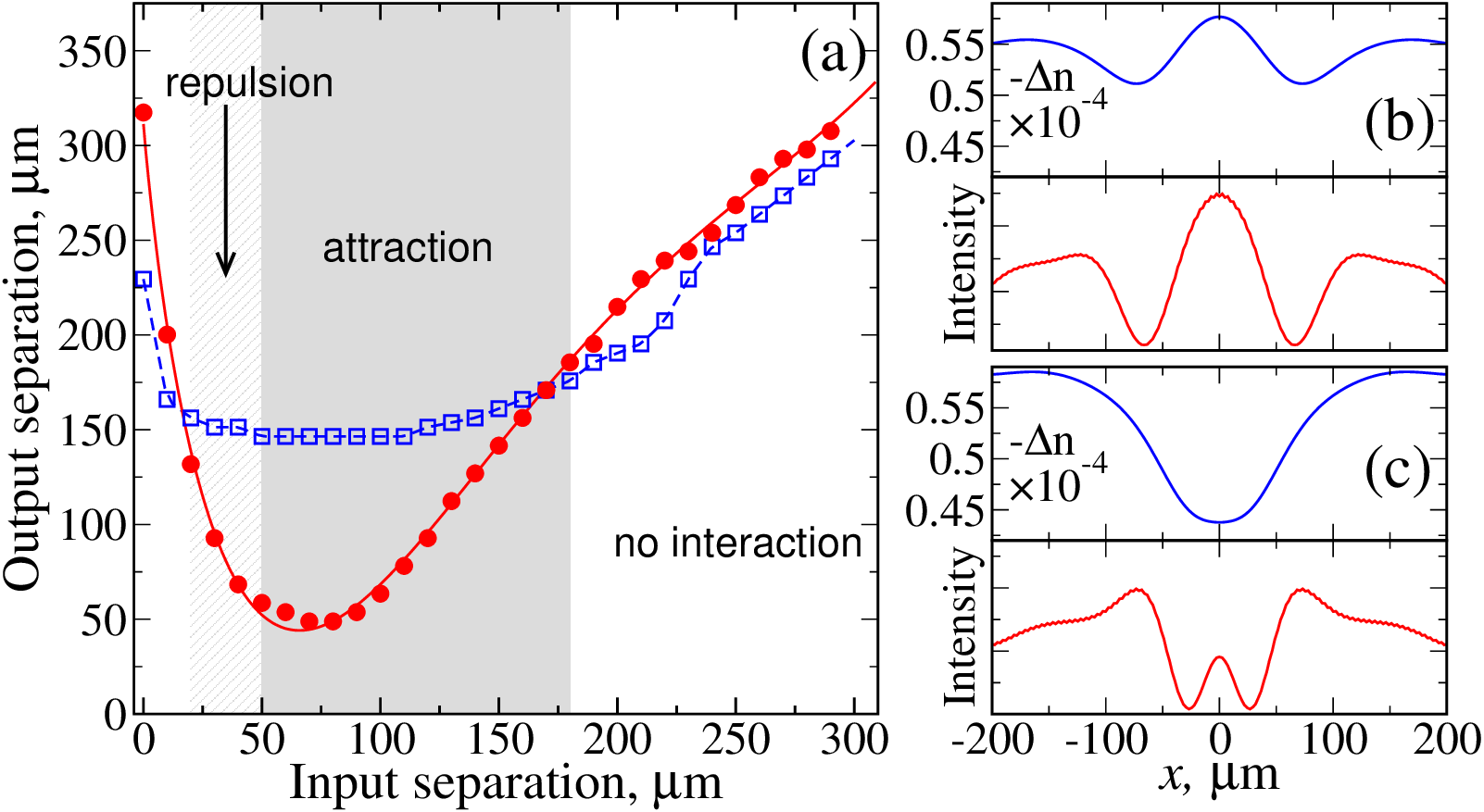}{fig2}{(color online) (a) Calculated output separation ($z$=50~mm) versus initial separation between two dark solitons for low ($I_0=0.01$~- squares) and high ($I_0=0.5$ - circles) intensity of the background beam. The different regimes of interaction are marked. (b,c)~Output intensity (bottom) and index (top) profiles for two different initial separations -- (b) repelling solitons (20~$\mu$m) and (c) attracting solitons (90~$\mu$m).}
%--------------------------------------------------------------------------

The spacing between the solitons obviously determines the shape of the resulting potential, and subsequently the strength of their interaction. In Fig.~\rpict{fig2}(a) we present the numerically determined separation between the two dark solitons, after propagation of 50~mm, as a function of their input spacing for low (linear regime~-- squares) and high intensity (nonlinear regime -- circles) of the background beam. For initial spacings larger than 180~$\mu$m the final separation is close to that of the linearly diffracting dark beams as the nonlocality does not contribute appreciably to the already weak interaction. The most interesting regions in Fig.~\rpict{fig2}(a) correspond to closely spaced solitons (20--180~$\mu$m). It is evident that in this region the separation between the solitons is a non-monotonic function of the initial spacing, being smaller than that in the linear regime. This behavior is a direct manifestation of the nonlocality induced {\em attractive forces}.
For closely spaced initial dark notches (20-50$\mu$m), the generated solitons strongly repel and their final separation is {\em larger} than the initial one. In the case of intermediate separations (50-180$\mu$m), the attractive forces can balance the natural repulsion and the solitons exhibit oscillating trajectories. Plots in Fig.~\rpict{fig2}(b,c) illustrate the output intensity and the refractive index profiles (which plays the role of confining potential) in those two cases. For the diverging solitons the potential has a form of two distinct wells separated by a barrier, whereas attraction occurs when the potential represents a single well.

%%%Experiment%%%%%
%--------------------------------------------------------------------------
\pict[0.95]{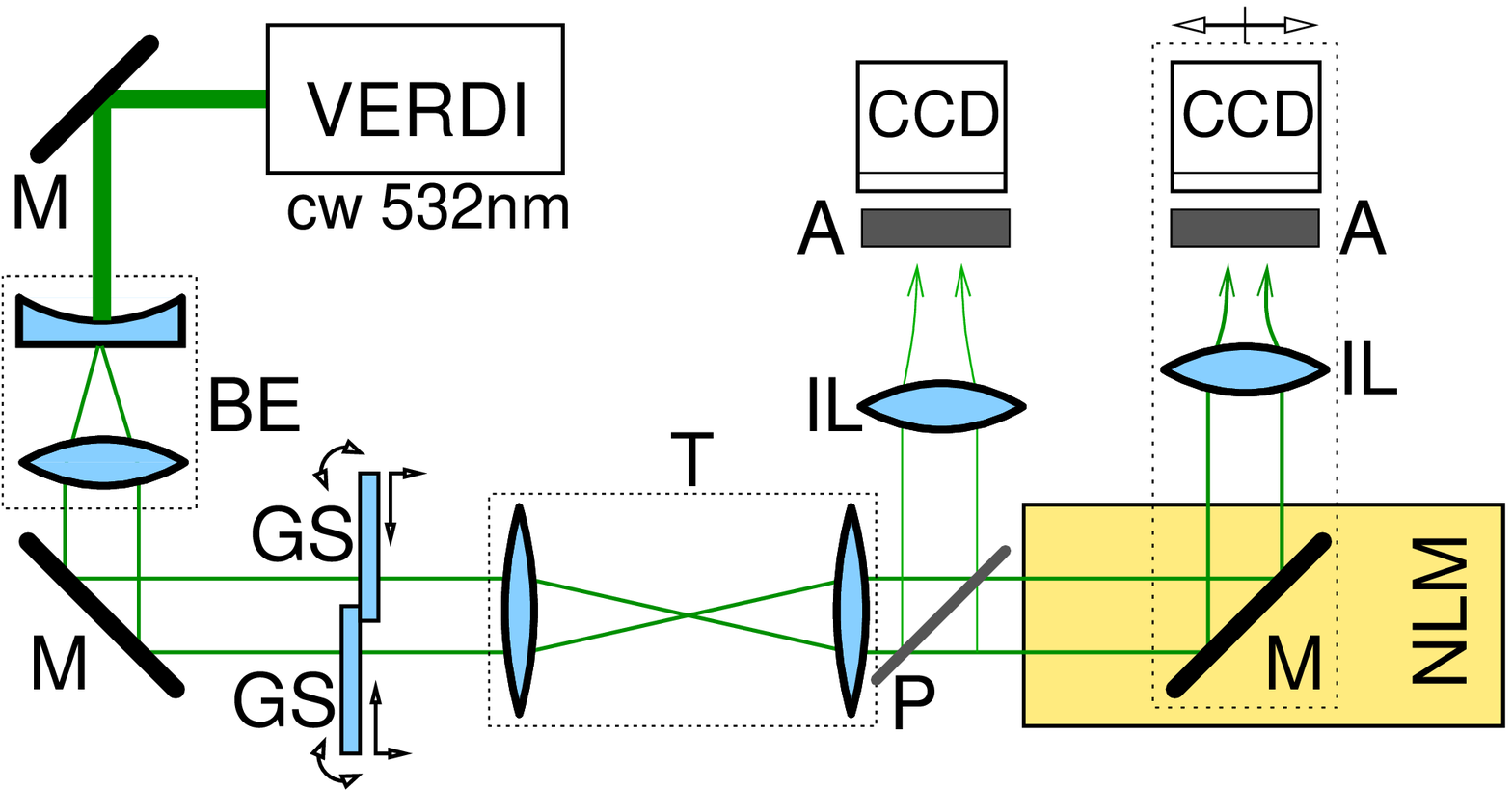}{setup}{(color online) Experimental setup: M - mirrors, GS - glass slides, BE - beam expander, T - imaging telescope, P - pellicle, IL~-~imaging lenses, A - attenuator, NLM - nonlinear medium, and CCD - cameras.}
%--------------------------------------------------------------------------

To investigate experimentally the interaction of dark solitons in a nonlocal nonlinear medium we used the experimental setup shown in Fig.~\rpict{setup}. A laser beam from a frequency doubled solid-state laser (Verdi-V, $\lambda=532$nm) was expanded by a system of lenses and passed through two closely overlapping microscope glass slides. The slides were subsequently imaged by a telescope onto the input face of a 50~mm long glass cell. The resulting beam diameter at the cell's input was FWHM=2.8~mm. The cell was filled with paraffin oil dyed with iodine. The iodine served as a weak absorber of the green light and its low concentration of 0.5~mg/l ensured that the total absorption in the cell was $\approx 10\%$. The density of paraffin oil decreases with increasing  temperature, thus resulting in a {\it self-defocusing nonlinear response}. The two glass slides were tilted at a small angle with respect to the beam and modified the phase structure of the beam such that it results in $\pi$ phase jumps in the beam at the position of the glass edge. The phase modulation also gives rise to amplitude modulation at the front face of the cell as seen in the inset of Fig.~\rpict{fig4}. The input and the output facets of the cell were imaged onto two CCD cameras by large numerical aperture lenses. The beam evolution along the propagation direction was traced by immersing a translatable mirror into the liquid and visualizing the beam profile at different distances~\cite{Neshev97}. Because of the finite size of this mirror, however, the first 19~mm of beam propagation could not be directly accessed.

The input beam phase profile was monitored by interference with a reference plane-wave [Fig.~\rpict{fig4}(top inset)]. Both phase jumps were set to $\pi$ within the accuracy of our interferometric measurement ($\sim$10\%). This accuracy was increased by monitoring the beam profile at the far field and setting the intensity in the dark notch to zero. By varying the transverse position of the slides we could change the initial separation between the generated dark notches. A typical intensity profile at a separation of 59~$\mu$m is shown in [Fig.~\rpict{fig4}(bottom inset)]. The width of each notch is $a\sim 18~\mu$m, which corresponds to $\sim$10 diffraction lengths of propagation in the nonlinear medium. The individual propagation in the self-defocusing medium results in formation of a ``black'' soliton of zero transverse velocity. The experimental conditions for dark soliton formation were determined by attaining saturation of the, so-called soliton constant, $I_0a^2,$ with input beam power~\cite{Dreischuh:1999-6111:PRE}. In our experiments this regime was reached at approximately 3~W of input power. All our measurements were performed at slightly higher power levels (3.5~W) to assure soliton regime in the presence of interaction and absorption.

%--------------------------------------------------------------------------
\pict[0.95]{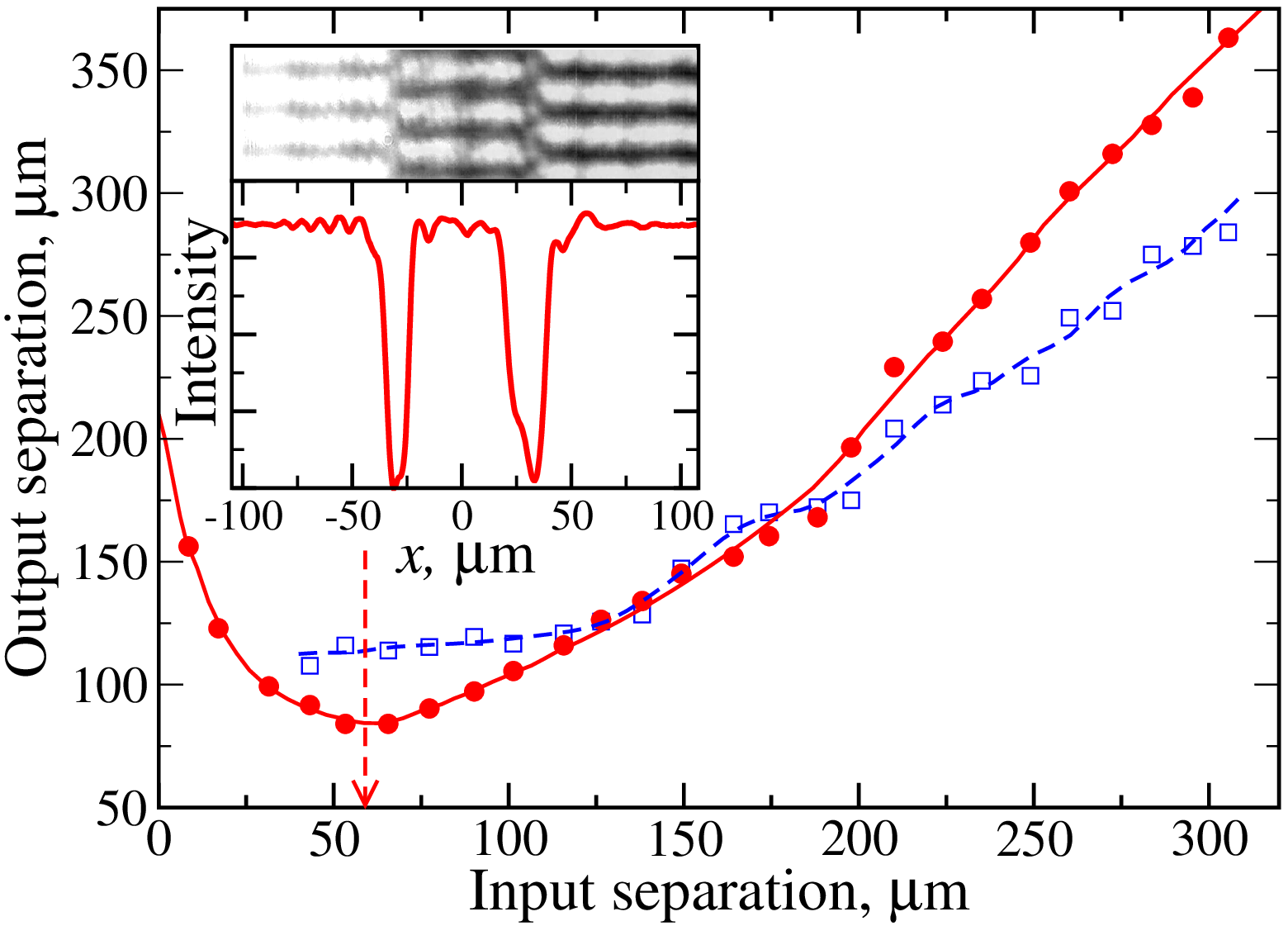}{fig4}{(color online) Measured output distance between two dark solitons as a function of their initial separation. Dots - nonlinear regime (3.5~W); Squares - linear propagation (10~mW). For small initial separations the output dark notches are practically not detectable in the linear regime. Inset -- interferogram reflecting the initial phase and intensity profile of solitons separated by 59~$\mu$m.}
%--------------------------------------------------------------------------

When placed sufficiently close, both dark solitons (with parallel initial trajectories) interact during propagation. This influences their output separation, which is compared to the separation at low power. In Fig.~\rpict{fig4} we depict the spacing between the dark solitons at the exit of the cell as a function of their initial separation. The dots represent nonlinear regime (3.5W), while squares correspond to linear propagation (10~mW). The measurements quite faithfully reproduced the theoretical predictions. For large initial spacings ($>180\mu m$) the solitons are weakly interacting, but their separation is affected by the background beam nonlinear broadening, resulting in increasing separation, which becomes larger than in the linear regime. For initial separations of 20-180$\mu$m the final spacing between the solitons is less than in the linear regime. This unambiguously indicates the presence of {\em attractive forces} counteracting the natural repulsion of solitons. Moreover, for input spacings of 115-200~$\mu$m the final separation is actually smaller than the initial one. It should be noted that the separations recorded in the nonlinear regime are detected with high accuracy ($\sim 1\mu m$) due to the localization of the dark notches. In the linear regime the determination of the central position of the dark notch is a subject of broadening and reshaping. The comparison of these two cases, however, gives us the correct information for the range of strong attraction.
%
%--------------------------------------------------------------------------
\pict{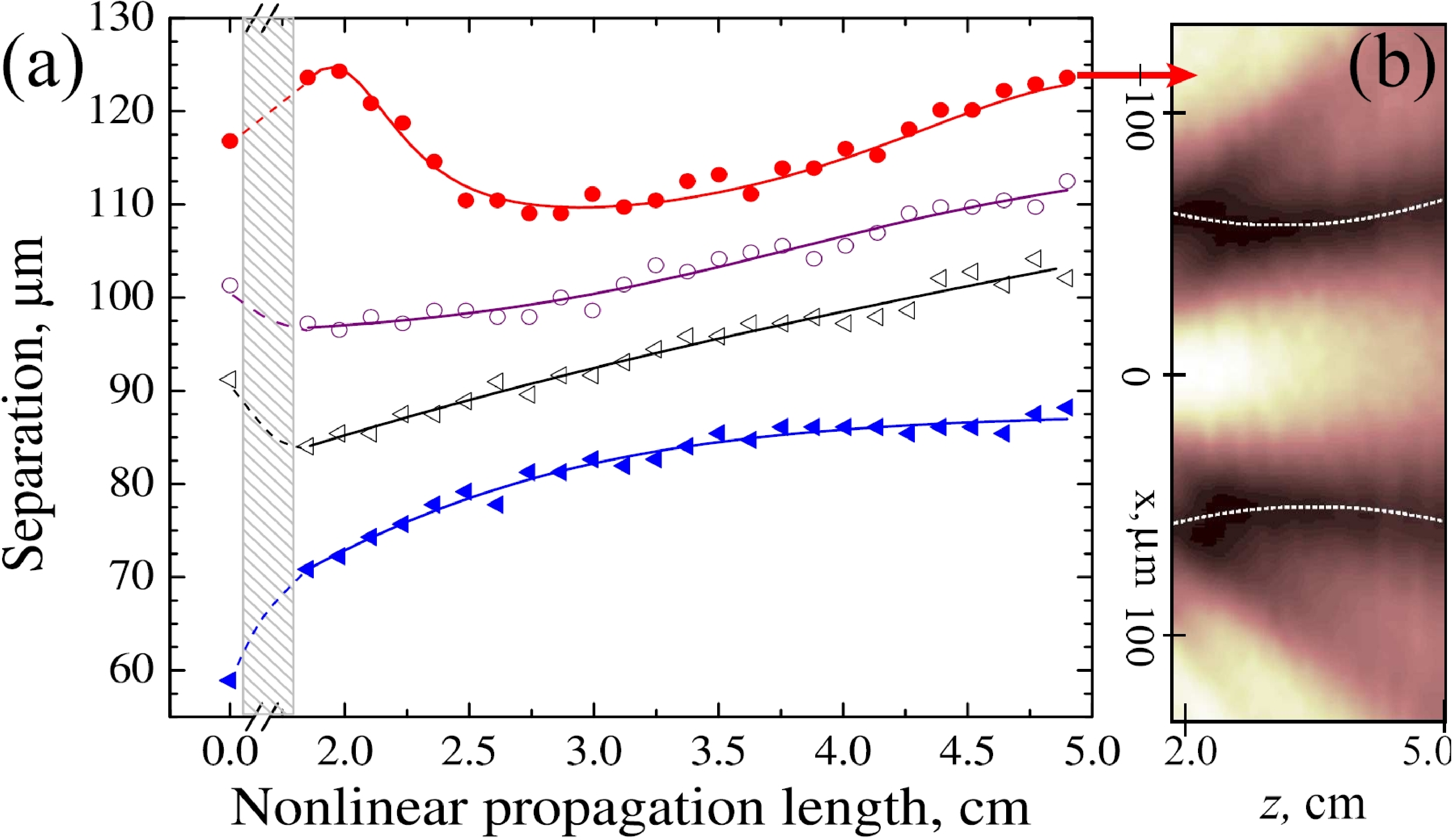}{fig5}{(color online) (a) Measured output separation of interacting nonlocal dark solitons in thermal medium as a function of the propagation distance for input separation of 117~$\mu$m - closed circles; 101~$\mu$m - open circles; 91~$\mu$m - open triangles; 59~$\mu$m~- closed triangles. (b) Measured trajectories of nonlocal solitons for initial separation of 117$\mu$m.}
%--------------------------------------------------------------------------

A clearer picture of soliton interaction can be obtained by following their trajectories inside the medium. This is illustrated in Fig.~\rpict{fig5} where we show the measured separation along the cell for initial spacings of 59, 91, 101, and 117~$\mu$m. For the separation of 59~$\mu$m the repulsive force is strong and can not be compensated by the nonlocality induced attraction. Therefore, for this separation the two dark solitons repel. The situation changes drastically for larger initial separations. At a certain distance the solitons actually come closer than their initial separation. This behavior cannot be directly observed for the initial separations of 91 and 101~$\mu$m since the decrease in soliton separation occurs at the initial non-accessible part of propagation. However, it is particularly visible for 117~$\mu$m separation.
% where the smallest separation due to the attractive forces is achieved for %28~mm of propagation.
The non-monotonous character of the trajectories is a direct proof of the interplay of repulsive and {\em nonlocality-mediated attractive forces} acting between the solitons. When the attraction dominates, solitons decrease their mutual separation until the repulsion prevails forcing them to move apart. The contour plot in Fig.~\rpict{fig5}(b) shows the experimentally obtained trajectories in this regime (separation of 117$~\mu$m). Dashed lines indicate location of the intensity minima. The inward bending of these trajectories is clearly visible.

In conclusion, we have shown, what we believe is the first experimental demonstration of the nonlocality mediated attraction of dark spatial solitons. Our experimental observations are in good agreement with direct numerical simulations. We believe, that our results may be
%as this type of soliton interaction is generic for nonlocal nonlinearity,  are
applicable to other physical systems exhibiting nonlocal nonlinear response.
%including Bose Einstein condensate, where spatial nonlocality is a result of a long-range interparticle interaction.

The authors thank Yu.~S. Kivshar and A.~A. Sukhorukov for valuable discussions. This work was supported by the Australian Research Council. A.D. acknowledges support by the NSF-Bulgaria Grant F1303/2003.

%\vspace{-5mm}

\end{document}